%% file: 0_main.tex
\documentclass[sigconf]{acmart}

\setcopyright{none}
\renewcommand\footnotetextcopyrightpermission[1]{}
\usepackage[most]{tcolorbox}
\usepackage{svg}  
\usepackage{enumitem}
\usepackage{booktabs}
\usepackage{subcaption}

\begin{document}

\title[Student Perceptions and Preferences Regarding AI-Generated Instructional Videos]{Student Perceptions and Preferences Regarding AI-Generated Instructional Videos in Computing Education}

\author{Esse Ciego}
\email{esse@ece.ufl.edu}
\orcid{0009-0006-7144-4452}
\affiliation{%
  \institution{University of Florida}
  \city{Gainesville}
  \state{Florida}
  \country{USA}
}

\author{Shubbhi Taneja}
\email{staneja@wpi.edu}
\orcid{0000-0002-2403-9407}
\affiliation{%
  \institution{Worcester Polytechnic Institute}
  \city{Worcester}
  \state{Massachusetts}
  \country{USA}
}

\author{Wilson Wong}
\email{wwong2@wpi.edu}
\orcid{}
\affiliation{%
  \institution{Worcester Polytechnic Institute}
  \city{Worcester}
  \state{Massachusetts}
  \country{USA}
}

\author{Amanpreet Kapoor}
\email{kapooramanpreet@ufl.edu}
\orcid{0000-0003-1340-8315}
\affiliation{%
  \institution{University of Florida}
  \city{Gainesville}
  \state{Florida}
  \country{USA}
}

\input{1_abstract}




\maketitle

\input{2_intro}
\input{3_related_work}

\input{4_methods}
\input{5_results}
\input{6_discussion}

\input{7_limitations}
\input{8_conclusion}

\bibliographystyle{ACM-Reference-Format}
\balance
\bibliography{8_references}

\end{document}

%% file: 1_abstract.tex
\begin{abstract}
Students differ in how they prefer to engage with learning resources, with some favoring textual materials and others visual or video-based content. Recent advances in generative AI have led CS education research to focus on text-based AI tools for developing learning resources. However, advances in AI video models and the rapid proliferation of AI video generation tools have made it possible for instructors to create high-quality personalized educational videos efficiently and cost-effectively. Understanding students' perceptions of AI-generated videos is thus critical for helping CS instructors know when and how to use them purposefully. To address this gap, we conducted a descriptive post-test survey study in which 170 computing students at two U.S. institutions watched three 3-minute AI videos on the Markdown markup language created with Knowlify. Students then completed a survey about their perceptions of the Markdown videos and their broader views on the use of AI-generated videos in education. Students rated the Markdown videos as high-quality, accurate, and usable, with nearly half unable to determine whether the videos were AI-generated. At the same time, students expressed limited comfort with the widespread adoption of AI videos in the classroom. They preferred AI videos for simple, supplemental, and visual use cases, while expressing concerns about lower-quality or inaccurate content, reduced instructor interaction, and diminished educational value.
\end{abstract}

%% file: 2_intro.tex
\section{Introduction}

% Esse:
As human-recorded instructional videos became a staple resource for  CS students to review material and supplement live lectures \cite{mcgowan_teaching_2016, settleEtAl-2011-lectureCapture}, AI-generated instructional videos (AI videos) have emerged as another promising way to enhance student learning at scale. Today's AI video models are producing more realistic and customizable videos \cite{liu_survey_2026} that instructors can integrate as lectures into their classes. These efforts have already started across higher education, including business \cite{vallis_student_2024}, medicine \cite{temsah_openais_2025}, and CS \cite{arkun-kocadere_video_2024}. Notably, students who watch AI videos have shown comparable performance \cite{arkun-kocadere_video_2024, netland_comparing_2025, leiker_generative_2023} and even stronger retention \cite{pi_influences_2022, xu_recorded_2025} than human-recorded videos. And with AI-generated videos being cheaper and more efficient to produce than professional-grade instructional videos \cite{pellas_influence_2023}, instructors stand to save significant time and effort.

Despite their promise, AI videos risk resistance from CS students. Students already tend to feel skeptical using LLMs for study and coding help, citing concerns with inaccuracies \cite{zastudil_generative_2023, keuning_students_2024, bouvier_rest_2025} and biases \cite{amoozadeh_trust_2024}. Similar concerns arose with use of AI videos in other disciplines, where students reported feeling less trust and naturalness compared to human instruction \cite{vallis_student_2024}. To our knowledge, only one study has examined CS students' perceptions of AI videos directly: Arkun et al. who found that students felt distracted and uncomfortable watching an AI avatar teaching gamification \cite{arkun-kocadere_video_2024}. Yet, no study has examined how students perceive AI-generated videos for learning programming concepts or identified the contexts in which students consider AI videos to be appropriate for learning. 

Our descriptive study addresses this gap by conducting a post-survey study across two U.S. universities to better understand students' perceptions of AI videos in computing education. As part of this study, 170 students watched three short AI videos on how to use the Markdown markup language. These videos were produced using a one-shot prompt in the Knowlify platform, which generates explainer-style videos from prompts and source materials \cite{knowlify_homepage}. Then, they completed a survey on their opinions of both the Markdown videos and AI videos more broadly. We found that while students perceived the Markdown videos as high quality and usable, they also expressed reservations about broader use of AI videos in courses. Additionally, they identified learning contexts in which such videos are appropriate. Our findings  contribute empirical evidence on students’ perceptions of AI videos within the computing education community and provide practical guidance for instructors on how to adopt and deploy AI videos effectively.

%% file: 3_related_work.tex
\section{Related Work}

% Will mention abbreviations in the introduction:
% AI-Generated Instructional Videos -> AI videos
% Human-Generated Instructional Videos -> Human videos

% Will mention learning performance being similar/comparable for AI videos vs. human videos:
    % Might be persuasive for instructors

 \subsection{AI Videos in Higher Education} 
 University students' perceptions of use of AI-generated videos in higher education are mixed. Students tend to describe AI videos as less personal and less human: Vallis et al. found students in a business informatics course missed the naturalness and spontaneity with human instructors \cite{vallis_student_2024}, while Xu et al. found students had lower motivation and trust in AI videos \cite{xu_recorded_2025} to learn English vocabulary words. Quality concerns have also surfaced. For instance, Weerakoon found university instructors observed audio-visual mismatches and inaccurate content in AI-generated course introduction videos \cite{weerakoon_enhancing_2024}. Yet, not all experiences were negative. Participants in Leiker et al.'s study reported that AI-generated videos met their expectations and improved their understanding \cite{leiker_generative_2023}, suggesting the effectiveness of AI videos may vary across course contexts and learner populations \cite{vallis_student_2024, waisberg_openais_2024}. Within computing education, to our knowledge, Arkun et al.’s work is the only study examining students’ perceptions of AI videos  \cite{arkun-kocadere_video_2024}. They found that students perceived AI-generated appearances and voices as distracting and uncanny, even when the videos used the same transcript as human-produced versions.  However, Arkun et al.'s study focused on HCI concepts rather than programming and was conducted in Turkish, leaving a gap our study addresses by examining student perceptions of AI videos for programming instruction in English.

\subsection{Perceptions of GenAI in CS Education} 
Students reported value in using generative AI for CS learning, including text-based LLMs (e.g., ChatGPT, Claude) to better understand concepts \cite{zastudil_generative_2023, amoozadeh_trust_2024} and to study for exams \cite{rogers_attitudes_2024}. Hou et al. further found students valued asking follow-up questions without fear of TA judgment and appreciated its use for rare error messages that online resources like Stack Overflow can miss \cite{hou_effects_2024}. However, students and instructors share significant concerns, especially around trust. Both point to inaccuracies \cite{zastudil_generative_2023, prather_robots_2023, keuning_students_2024, bouvier_rest_2025} and biases in AI algorithms \cite{amoozadeh_trust_2024, prather_robots_2023}. Programming experience is a key moderating factor on trust: interviews with students \cite{hou_effects_2024, zastudil_generative_2023}, a survey from upper-division instructors \cite{bouvier_rest_2025}, and observational studies \cite{pratherEtAl-2023-copilotNoviceProgrammers} suggest that while experienced students can critically evaluate and filter GenAI responses, novice students over-rely on AI outputs in ways that undermine learning. Prather et al. also caution against GenAI use among novices, based on a working paper surveying 57 instructors across 12 countries \cite{prather_robots_2023}. Yet, most of this work focuses on text-based interactions with AI, leaving open whether these perceptions extend to videos generated using GenAI.

%% file: 4_methods.tex
\section{Methods\protect\footnotemark} 
\footnotetext{Sections of this paper were refined using GenAI tools (ChatGPT and Claude) to format tables \& figures, improve grammar, and clarity.}

\subsection{Study Design and Research Question}
To understand students’ perceptions and preferences for AI videos, we designed a post-test survey study in which students at two institutions watched three short AI-generated videos on learning how to use Markdown for GitHub elements such as README files. Students then completed a post-survey that measured their (a) perceptions of the videos used in the study, and (b) opinions on using AI-generated videos for CS courses including avenues appropriate for the videos and concerns about the use of such videos en masse. Through this study, we aim to answer the following research questions (RQs): 

\begin{itemize}
\item[RQ1.] What are computing students’ \emph{perceptions} of AI-generated instructional videos?

\item[RQ2.] In what instructional \emph{contexts} do students view AI-generated videos as \emph{appropriate} for learning in computing courses?

\item[RQ3.] What \emph{concerns} do students raise about the broader use of AI-generated videos in computing education?

\end{itemize}

% https://drive.google.com/drive/folders/1_gU9dHQYtBYxWbF1PmSHO6hXy0aSP2aM

\subsection{Research Sites and Participant Recruitment}
 Students were recruited from upper-level CS courses at two universities in the US. At the first institution, a large public university, students were recruited from a Data Structures and Algorithms (DSA) course. Admission to the university is competitive, and the course is required for computing majors. While Markdown is not explicitly covered in DSA course lectures or other courses that precede DSA, our CS1 and DSA courses have labs that use GitHub or a team project that expects students to use GitHub. Knowledge of Markdown for these labs and projects is expected to be learned independently, and textual resources are provided to students within these projects which they can refer to.

 At the second institution, a competitive polytechnic university, students were recruited from a Software Engineering course. Although not required, nearly all CS majors take this course as well as a few robotics engineering majors. This course includes a substantial team project that requires students to use GitHub and be able to create and modify README files formatted using Markdown. Students are expected to learn the language independently from the brief tutorial provided by GitHub on their website (such as \cite{github_basic_formatting}).

 Our study was approved by the local Institutional Review Board at both institutions. Students were offered a small amount of extra credit (1\% of their course grade) for participating. They were given the option to complete the study while excluding their data from analysis, or to complete an alternative assignment requiring similar effort for the same credit. 
 
 \subsection{Sample and Familiarity with Markdown}
  Of the 475 students enrolled in the respective courses at institutions 1 and 2, 215 students responded to the survey (Total Response rate: 45\%). Of these responses, 45 were excluded: 20 students who requested their data be excluded for research, 12 incomplete submissions, 2 duplicate submissions and 11 students who failed an attention check in the survey (where they were asked to mark ``Somewhat agree'' to an option). Thus, our analysis corpus consists of 170 consenting students (N=170, $N_{i_1}=143$ and $N_{i_2}=27$).

\begin{table}[ht]
\centering
\small
\caption{Self-reported Markdown familiarity by institution}
\label{tab:markdown-familiarity}
\scalebox{1}{
\begin{tabular}{lccccc}
\toprule
Institution & Not at all & Slightly & Moderately & Very & Extremely \\
\midrule
1 ($n=143$) & 37 (26\%) & 55 (38\%) & 36 (25\%) & 13 (9\%) & 2 (1\%) \\
2 ($n=27$)  & 4 (15\%)  & 12 (44\%) & 8 (30\%)  & 3 (11\%) & 0 (0\%) \\
\bottomrule
\end{tabular}
}
\par\smallskip
\footnotesize\textit{Note.} Values are $n$ (\%); percentages may not sum to 100 due to rounding.
\end{table}

The familiarity with Markdown was comparable across the two institutions in our sample (Table~\ref{tab:markdown-familiarity}). At both sites, most students reported little or no prior familiarity: 64\% of Institution 1 students and 59\% of Institution 2 students were not at all or only slightly familiar. Both a chi-square test of independence, $\chi^2(4)=2.01$, \textit{p}=0.73, and a Fisher's exact test (\textit{p}=0.74) indicated no significant association between institution and familiarity level, indicating that the students across the two institutions are statistically comparable.

\subsection{AI-generated videos}
\textbf{AI video creation platform.} The three AI-generated videos were created using the Knowlify platform \cite{knowlify_homepage}. Knowlify is a conversational AI video generation system that enables users to create explainer-style videos from natural language prompts and uploaded source documents. Users begin by providing a prompt and optionally uploading supporting materials, and selecting configurable options such as video length and narrator voice. The system then automatically generates a narration script, scene-level storyboard, and voiceovers, all of which can be reviewed and edited before the final video is produced.

\vspace{4px}
\noindent
\textbf{Logistics of video creation.} In our study, we uploaded GitHub's official Markdown formatting documentation \cite{github_basic_formatting} as a PDF to Knowlify to generate short 3-minute videos, a video length students prefer \cite{guo_how_2014}. Creating each video required 8-11 minutes for planning, reviewing, and configuring the video. The video generation process took 3 minutes and 48 seconds on average per video and consumed an average of 5,200 Knowlify credits, corresponding to \$52 under the paid plan pricing model of 1 cent per credit. Overall, creating a single 3-minute video took about \$50-60 and 15 minutes of design and system generation time, in addition to 10 minutes of human time for verification. Thus, total time taken was at most 25 minutes for each 3 minute video. All videos can be accessed here \cite{video_repo}.

\begin{figure}[ht]
    \centering
    \includegraphics[width=0.75\linewidth]{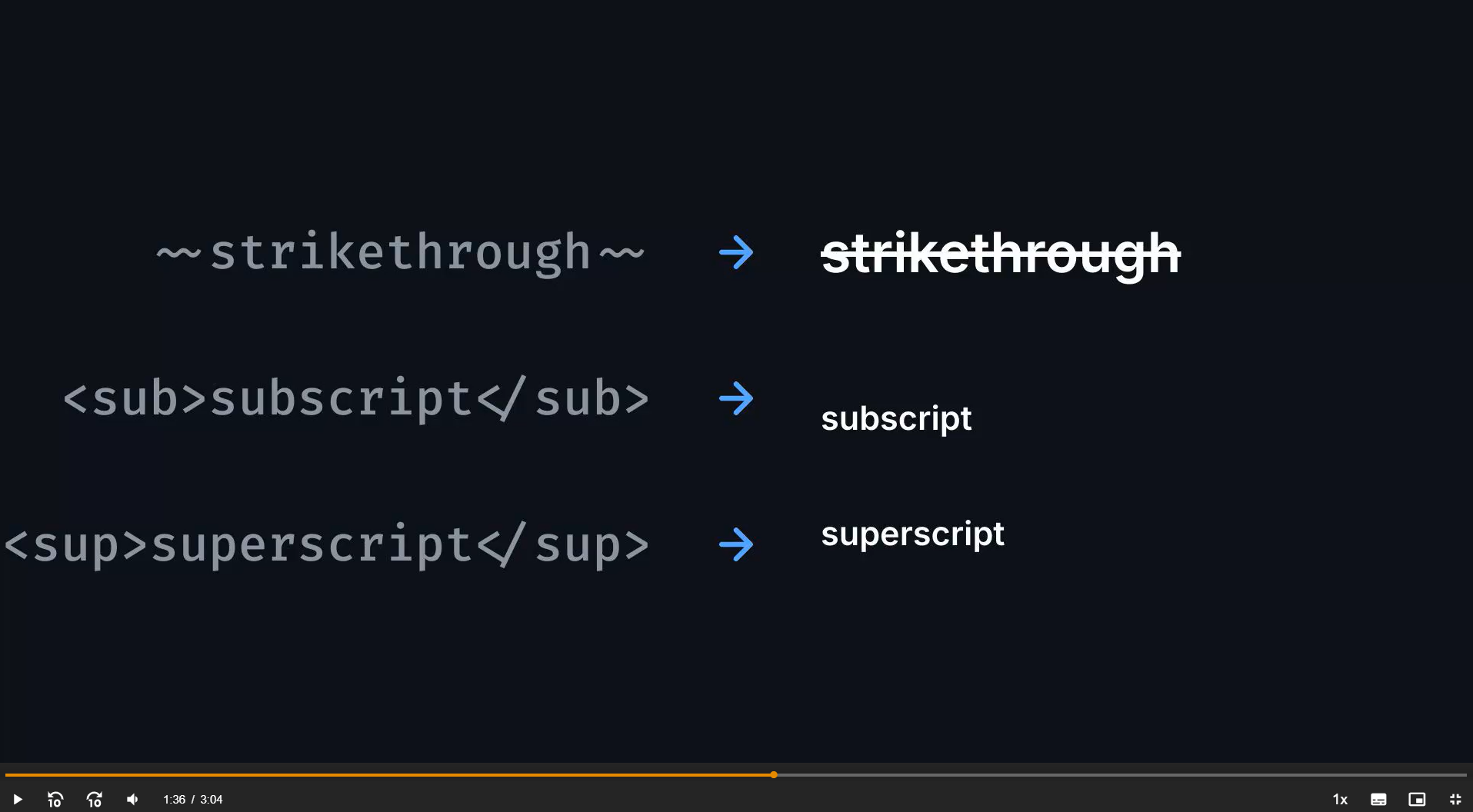}
    \caption{Video Frame from AI-generated Video 1: Intro to Markdown Markup Language - Text Formatting Blocks}
    \label{fig:md}
\end{figure}

\vspace{4px}
\noindent
\textbf{Video delivery and content.} 
The three videos totaled 9 minutes and 31 seconds and were hosted on a university-subscribed platform that required institutional login and provided detailed logs of student viewing behavior. The videos consisted of animations and visual walkthroughs but did not contain AI-generated avatars, deep-fakes, or talking heads (see Figure \ref{fig:md}). This design choice was intentional to minimize potential confounding effects of social presence \cite{social_presence}. Video 1, titled \textit{``Intro to Markdown Markup Language: Text Formatting Blocks''}, had a duration of 3 minutes and 4 seconds. This video covered the importance of Markdown and basics of formatting including headings and text emphasis features such as bold, italics, strikethrough, and subscripts.  Video 2, titled \textit{``Linking in Markdown''}, had a duration of 3 minutes and 15 seconds. This video covered the creation of hyperlinks in Markdown, and four types of links: inline links, section links, custom links, and relative links. Video 3, titled \textit{``Lists, Images and Alerts in Markdown''}, had a duration of 3 minutes and 12 seconds. This video covered topics such as how to create ordered, unordered, nested, and tasks lists as well as how to add images, footnotes and alert block quotes on GitHub using Markdown.

\subsection{Data Collection}
Students completed a post-survey on Qualtrics after watching the three videos. Links to the videos were provided both in a Canvas assignment and at the beginning of the survey. The post-survey included informed consent and 24 questions, 19 of which we analyzed in this paper based on their relevance to the research questions. These questions belonged to four survey sections:
\begin{enumerate}
    \item confirmatory item on watching the videos, and familiarity with Markdown (2 questions)
    \item a knowledge test with factual and interpretive items based on the video content on Markdown (5 multiple choice questions with four options)
    \item Likert-scale items assessing perceptions of Markdown videos (6 questions), an attention check (1 question), and perceptions of AI video use in future (3 questions)
    \item two open-ended questions on use of AI videos in education
\end{enumerate}

To address RQ1, students were not told \emph{apriori} that the videos were AI-generated to avoid biasing their evaluations \cite{scharowski_certification_2023, yin_ai_2024}. Instead, a post-survey item asked them to rate whether they could tell it was AI-generated after they had answered their perception ratings on the Markdown videos. On average, students completed the survey in 15.5 minutes.

% An example of a factual question was: \textit{How do you make text bold in Markdown?} Options: \textit{(1) **text**; (2) -{}-text-{}-; (3) ==text==; (4) ++text++}. An example of an interpretive question was : \textit{Given the following Markdown, what will the rendered output look like? **This text is \_extremely\_ important**}. Options: \textit{  (1) All text bold, with ``extremely'' also italicized; (2) All text italicized, with ``extremely'' also bold; (3) Only ``extremely'' is bold and italic; the rest is plain; (4)  The underscores are displayed literally because they are inside bold markers}.

\subsection{Data Analysis}
Survey data were analyzed using quantitative and qualitative techniques. We present descriptive statistics of the multiple choice and Likert scale questions (coding \textit{Strongly agree}: 5 and \textit{Strongly disagree}: 1) for RQ1. For RQ2 and RQ3, we analyze open-ended responses using inductive thematic analysis \cite{braun2006using}. The qualitative data were coded into primary codes, which were abstracted into broader themes. To ensure the reliability of the coding scheme, the first and last authors reviewed the codebook and discussed the identified themes. In cases of disagreement, the terminology and definitions were refined through an iterative consensus-building process. We also present a frequency analysis on the resulting codes and themes.

%% file: 5_results.tex
\section{Findings}

\begin{figure*}
\centering
\includegraphics[width=1.0\linewidth]{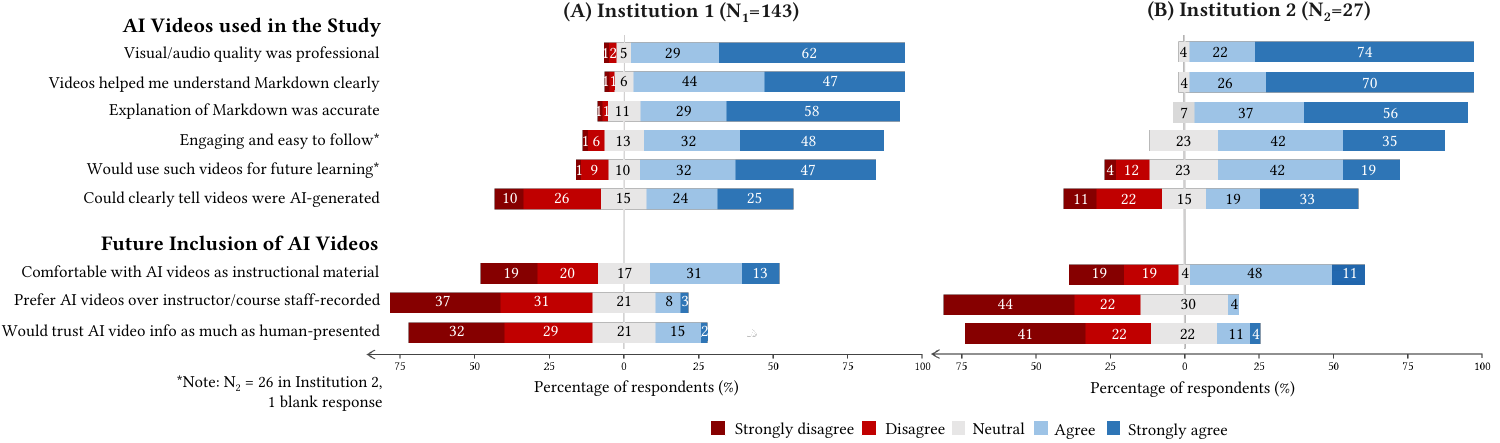}
\caption{Distribution of student perceptions (N=170) for AI videos used in the study and future inclusion of AI Videos in coursework. Bars show the \% responses at each Likert level, centered on neutral.}
\label{fig:video-perceptions}
\end{figure*}

%  % at (A) Institution 1 ($N_{i1}=143$) and (B) Institution 2 ($N_{i2}=27$)

% \begin{figure*}[t]
% \centering
% \begin{subfigure}{0.49\textwidth}
%   \centering
%   \includegraphics[width=\linewidth]{images/ai_video_perceptions_combined_1.pdf}
%   \caption{Institution 1 ($N=143$)}
%   \label{fig:video-perceptions-i1}
% \end{subfigure}\hfill
% \begin{subfigure}{0.49\textwidth}
%   \centering
%   \includegraphics[width=\linewidth]{images/ai_video_perceptions_combined_2.pdf}
%   \caption{Institution 2 ($N=27$)}
%   \label{fig:video-perceptions-i2}
% \end{subfigure}
% \caption{Distribution of student perceptions for AI videos used in the
% study and future inclusion of AI videos in coursework. Bars show the \%
% of responses at each Likert level, centered on neutral.}
% \label{fig:video-perceptions}
% \end{figure*}

\subsection{Study Context: AI Videos Used in this Study}

\textbf{Video watching behavior.} Our survey included a self-report item asking, \textit{``I watched the Markdown markup language videos assigned to me''}: with response options indicating whether students watched all, some, or none of the videos. At institution 1 ($N_{i_1}=143$), 92\% students self-reported watching all of the videos, while 8\% reported watching some of the videos or only partially viewing them. At institution 2 ($N_{i_2}=27$), 89\% students self-reported watching all of the videos, while 11\% reported watching some of the videos.  None of the students reported they did not watch any of the videos. Additionally, the average watch coverage (how much proportion of the video did a student play) was 77-81\% for the three videos in the logs (Video 1: 79\%, Video 2: 77\% and Video 3: 81\%) at institution 1 and 76-80\% (Video 1: 76\%, Video 2: 78\% and Video 3: 80\%) at institution 2 verifying student self-reported numbers. This suggests that nearly all students in the sample engaged with the video content.  

\vspace{5px}
\noindent
\textbf{Learning outcomes associated with video use.} The average aggregate test score across the five Markdown quiz questions (possible range: 0-5) at institution 1 was $\mu_{i_1} = 4.3$ (min: 1; max: 5; median: 5; $\sigma = 0.88$) and at institution 2 was $\mu_{i_2} = 4.3$ (min: 3; max: 5; median: 4; $\sigma = 0.62$). This suggests that students learned fairly well about Markdown from the AI-generated videos.

\subsection{Perceptions of AI Videos (RQ1)}

\noindent
\textbf{Markdown videos used in our study.}
Students rated the Markdown AI videos favorably across all quality dimensions in both institutions (see Figure \ref{fig:video-perceptions}). A large majority of the 170 students agreed or strongly agreed that the videos were professionally produced ($\mu=4.52$, $\sigma=0.77$; 92\% agreement), helped them understand Markdown ($\mu=4.39$, $\sigma=0.76$; 92\% agreement), and explained Markdown accurately ($\mu=4.43$, $\sigma=0.78$; 88\% agreement). Ratings of engagement ($\mu=4.20$, $\sigma=0.90$; 80\% agreement) and willingness to use similar videos for future learning ($\mu=4.06$, $\sigma=1.04$; 76\% agreement) were slightly
lower but still strongly positive. In contrast, detectability was the only item without consensus: just  half of students (50\%) agreed they could clearly tell the videos were AI-generated, while 35\% disagreed and 15\% were neutral ($\mu=3.31$, $\sigma=1.36$) --- the highest variance of any item. 

% \begin{tcolorbox}[
%   enhanced, frame hidden,
%   borderline west={2pt}{0pt}{blue!60!black},
%   colback=gray!8,
%   left=10pt, right=8pt, top=6pt, bottom=6pt,
%   sharp corners
% ]
% \textbf{Key finding: Perceptions of used AI Videos.} Together, these results indicate that students watched the AI-generated videos, scored well in the post-assessment and perceived them to be high-quality, accurate, and usable instructional material, even though roughly half of the students could not reliably identify them as AI-generated.
% \end{tcolorbox}

\vspace{5px}
\noindent
\textbf{Future inclusion of AI videos in courses.} Despite rating the study videos highly, students were considerably more restrained about the broader inclusion of AI-generated videos in their coursework (see Figure \ref{fig:video-perceptions}). Comfort with using AI-generated videos ``\textit{as instructional material in my courses}'' was nearly evenly split, with 46\% agreeing and 39\% disagreeing ($\mu=3.00$, $\sigma=1.34$). However, two items drew clear opposition: 68\% of 170 students disagreed that they would prefer AI videos over videos recorded by their instructor or course staff ($\mu=2.06$, $\sigma=1.06$; only 10\% agreed), and 62\% disagreed that they would trust information in an AI video as much as in a human-presented one ($\mu=2.24$, $\sigma=1.14$; 17\% agreed).

% \begin{tcolorbox}[
%   enhanced, frame hidden,
%   borderline west={2pt}{0pt}{blue!60!black},
%   colback=gray!8,
%   left=10pt, right=8pt, top=6pt, bottom=6pt,
%   sharp corners
% ]
% \textbf{Key finding: Perceptions of using AI Videos in Future Courses.} Although students favorably rated the AI videos used in this study, the results suggest they had limited comfort for using AI videos broadly in coursework. Their reservations stem less from perceived production quality or clarity than from a preference for human instructors and lower trust in AI-presented information.
% \end{tcolorbox}

\subsection{Appropriate avenues for AI Videos (RQ2)}
We coded 163 non-blank responses to an open-ended question \textit{"Describe one educational scenario where AI-generated videos would work well and one where they would not. Explain your reasoning for each."} into 14 unique codes abstracted into 5 themes. Three students (2\%) believed there were no use-cases where AI would work well.

\subsubsection{Appropriate avenues}
\subsubsection*{\textbf{Theme 1: Topic Match}}
In this theme, students described AI videos were well suited to learn certain topics, particularly when the content was simple or could be easily visualized.

\vspace{4px}
\textbf{Simple topics} (n=44, 27\%)\textbf{:} Students identified foundational content (e.g., vocabulary, basic syntax, and logical processes) as suitable for AI videos. Some mentioned Markdown video's simplicity as a good use of this format, as the videos are \textit{``simple, rigid, and well-documented by online resources''} (R160). 

\vspace{4px}
\textbf{Topics with visualizations} (n=24, 15\%)\textbf{:} Students saw topics with images or graphs, complex processes, demo of program behavior, and walk-through tutorials as good candidates for AI videos. For example, R36 said,\textit{``AI-generated videos work extremely well for explaining data structures like Linked Lists \& Trees since they can add elements with an animation, show what traversal would look like.''}

\subsubsection*{\textbf{Theme 2: Delivery Match}}
Besides content, students felt AI videos were well suited to specific delivery tasks, particularly clarifying, summarizing, replacing, or personalizing content.

\vspace{4px}
\textbf{Topic summaries} (n=20, 12\%)\textbf{:} Students believed AI videos would be useful for high-level overviews and for  
% introducing new topics and 
reviewing exam material. For instance, R100 explained that \textit{``Because it does not have to be extremely thorough, I would trust AI generated videos to give a recap, highlighting the most important details''.}

\vspace{4px}
\textbf{Supplemental learning material} (n=20, 12\%)\textbf{:} Students saw AI videos useful for clarifying difficult concepts and deepening understanding. As R1 described, AI videos can add \textit{``additional info described in a slightly different way to reinforce concepts.''}

\vspace{4px}
\noindent
\textbf{Replace written learning material} (n=11, 7\%)\textbf{:} Students saw potential in AI videos to replace online resources by breaking down dense text. As R38 explained, \textit{"[W]atching a short video on something is a lot less daunting than reading a tedious article"}.

\vspace{4px}
\textbf{Tailor learning} (n=8, 6\%)\textbf{:} Students recognized AI videos' potential to personalize learning through tailored content, targeted feedback, and multilingual accessibility. R123 believed AI videos can help struggling students \textit{``catch up in learning the fundamental skills while other students can pursue other academic goals''}.

\vspace{4px}
\noindent
\textbf{Improve presentation clarity} (n=4, 2\%)\textbf{:} Students also saw value in AI videos improve instructors' audio and visual presentation quality. R50 gave an example of a professor using AI to voiceover their own presentation to overcome a language barrier.

\subsubsection*{\textbf{Theme 3: Instructor Capacity Gap}}
Beyond topic and delivery considerations, students saw AI videos as a way to address gaps in instructor availability or institutional resources.

\vspace{4px}
\textbf{Lack proper instructors} (n=6, 4\%)\textbf{:} Students acknowledged that AI videos could serve as instructors in settings without proper funding or learning resources. R80 envisioned these facilities using AI videos to teach courses \textit{``they would not have been able to teach with a human teacher''} which they believed \textit{``would help bridge the gap for people who wouldn't have gotten that education otherwise''}.

\vspace{4px}
\textbf{Reduce instructors' workload} (n=6, 4\%)\textbf{:} Students also saw AI videos freeing up time for instructors to focus on more complex work, grading, and course content. As R116 said, \textit{``[AI Videos] could work well for quickly creating the material for students, this would also provide the teacher more time for the students or other priorities.''}

\subsubsection{Inappropriate avenues}

\subsubsection*{\textbf{Theme 1: Topic Mismatch}}
For this theme, students felt AI videos were poorly suited for certain topics, particularly when the topic required nuance, subjectivity, or hands-on engagement.

\vspace{4px}
\textbf{Complex topics} (n=50, 31\%)\textbf{:} Students believed AI videos were unsuitable for topics that were advanced, nuanced, abstract, difficult, or had complex diagrams. As R98 specified, \textit{``for more important concepts, like graphs and trees, it may be hard to use AI videos because students are going to need more elaboration and breakdown.''}

\vspace{4px}
\textbf{Subjective topics} (n=22, 14\%)\textbf{:} Students believed humanity-based (e.g. philosophy, politics, art) and communication-based (e.g. public speaking) courses were unsuitable for AI videos. Topics with diverse experiences or backgrounds were also unfit as AI might perpetuate biases.  As R108 said, \textit{``AI would not work well in a scenario where there is no correct answer or deterministic process.''}

\vspace{4px}
\textbf{Niche topics} (n=6, 4\%)\textbf{:} Students highlighted specialized topics with limited online information as challenging for AI videos. R63 said, for example, \textit{``AI may not get the right sign language gestures right because they are notoriously bad with hands.''}

\subsubsection*{\textbf{Theme 2: Delivery mismatch}}
In addition to content limitations, students identified specific learning contexts where AI videos would not be effective, such as replacing lectures.

\vspace{4px}
\textbf{Replace all instructors' lectures} (n=26, 16\%)\textbf{:} Students argued that AI videos should not replace instructor-led in-person lectures, online recorded lectures, or exam reviews.

\vspace{4px}
\textbf{Interactive learning} (n=14, 9\%)\textbf{:} Students felt AI videos were poorly suited for hands-on learning like debugging code or designing algorithms, which require human instruction. Discussion-based classes (e.g. computing ethics seminar) might also pose problems. As R144 explained, discussions \textit{``require dynamic back-and-forth perspective sharing and real-time responses to student viewpoints''.}

\subsection{Concerns with AI Videos (RQ3)}
We coded 163 non-blank responses to an open-ended question -- \textit{``What concerns, if any, do you have about using AI-generated videos in education?''} using 10 unique codes which were abstracted into 3 themes. There were nine students (6\%) who reported no concerns.

\subsubsection*{\textbf{Theme 1: Low Content Quality}}
For this theme, students feared AI videos would deliver lower-quality lectures than instructors, affecting lessons' depth, engagement, and trustworthiness.

\vspace{4px}
\textbf{Inaccurate information within the video} (n=94, 58\%)\textbf{:} Many students noted AI can hallucinate, leading them to trust their instructors over AI videos. As R25 explained, \textit{``although professors *can* make mistakes, my experience of AI is that it *will*''}.

\vspace{4px}
\textbf{Lack of depth in video content} (n=22, 13\%)\textbf{:} Students worried AI videos might oversimplify concepts, skip steps, or leave out information needed for exams. As R134 said, AI videos \textit{``may not be able to explain the content very well because they may just say what something is and not explain it in human words''}.

\vspace{4px}
\textbf{Lack engagement} (n=21, 13\%)\textbf{:} Students worried AI videos would struggle to keep their attention. They noted presentation issues, including monotonous tone, robotic or dull delivery, and pronunciation errors. As R149 reported, \textit{`Sometimes the [Markdown] videos all sound the same and I tune them out after a while''}.

\vspace{4px}
\textbf{Lack instructors' insight and personal experiences} (n=9, 6\%)\textbf{:} Students felt AI videos lack the personal anecdotes and faculty insight needed to explain concepts understandably. As R4 explained, \textit{``having that personal instruction makes learning more enjoyable in my opinion and keeps things interesting''}.

\vspace{4px}
\textbf{Lack thorough fact-checking by an educator} (n=7, 4\%)\textbf{:} Students worried that videos would not be thoroughly reviewed for accuracy. Factors such as high production quality might \textit{``lull a false sense of security''} (R94), making errors easier to miss. Additionally, longer video lengths raised concerns that \textit{``instructors providing the videos did not watch it all the way through''} (R46).

\subsubsection*{\textbf{Theme 2: Disconnect Between Students and Educators}}
Beyond content, students anticipated AI videos would fundamentally change how they interact with instructors or TAs.

\vspace{4px}
\textbf{Fewer interactions with educators} (n=16, 10\%)\textbf{:} Students worried AI videos might limit face-to-face engagement, encourage remote learning, or replace teachers entirely. As R22 expressed that if the videos are used broadly, \textit{``it could cause more of a disconnect between the professor and students, and the students might start relying more on AI even more''}.

\vspace{4px}
\textbf{Lack ability to ask follow-up questions} (n=6, 4\%)\textbf{:} Students noted that AI videos limit their ability to get clarification. As R43 explained, \textit{``you can't ask follow-up questions in real time or get clarification tailored to your confusion''}.  

\subsubsection*{\textbf{Theme 3: Erode Educational and Societal Value}}
Beyond learning concerns, students worried how AI videos affect worth of college education and society at large.

\vspace{4px}
\textbf{Waste of college tuition} (n=17, 10\%)\textbf{:} Students felt AI videos undermine college's value proposition, making expensive tuition seem unjustified. As R146 commented, \textit{``people are not paying a fortune to get a degree from watching AI generated videos''}.

\vspace{4px}
\textbf{Broader societal impacts} (n=9, 6\%)\textbf{:} Students raised societal concerns (e.g. lower instructor pay or environmental impact of creating AI videos). As R111 noted, \textit{``using AI on a large scale is prohibitively wasteful in terms of data, computing power, and water''}.

\vspace{4px}
\textbf{Decrease instructors' effort in teaching} (n=5, 3\%)\textbf{:} Students expressed concern about instructors will not care what students are taught and use poor quality videos. As R19 explained, teachers risked \textit{``becoming lazy or complacent and just offloading all their teaching work to a suboptimal AI model''}.

% \begin{tcolorbox}[
%   enhanced, frame hidden,
%   borderline west={2pt}{0pt}{blue!60!black},
%   colback=gray!8,
%   left=10pt, right=8pt, top=6pt, bottom=6pt,
%   sharp corners
% ]
% \textbf{Main finding: Concerns of AI Videos.} Students' concerns about AI videos spanned \textbf{cognitive} (low content quality, especially inaccuracy), \textbf{situated} (disconnect from instructors and TAs), and \textbf{critical} (erosion of educational and societal value) dimensions. Inaccuracy was the most dominant concern, raised by over half of respondents.
% \end{tcolorbox}

%% file: 6_discussion.tex
\section{Discussion}
% NOTE: Each first sentence of each paragraph captures the main finding of quantitative data, and the rest of the paragraph supports the finding by talking about the qualitative data + literature

Students responded positively to the AI-generated Markdown videos, rating them highly across quality, helpfulness, and engagement. We propose three explanations for these results. First, the videos were short and simple, covering basic syntax that students saw as well suited for AI videos. This aligns with prior work showing students prefer more concise instructional videos for computing \cite{herala2017experiences, mcgowan_teaching_2016, scott_educational_2017}. Second, the videos did not replace instructor-led lectures, a concern many students had. Third, students rated production quality highly, unlike earlier work where AI avatars provoked discomfort \cite{arkun-kocadere_video_2024}. The fact that half the students could not reliably identify the videos as AI-generated suggests the increasing realism of current AI video generation technology. Our videos did not feature an AI avatar and benefited from new advances in the technology, likely reducing the uncanniness. Thus, our findings resemble Leiker et al. and Vallis et al. which showed general receptiveness to AI videos \cite{leiker_generative_2023, vallis_student_2024}.

Despite positive ratings, students expressed distrust toward AI videos if used as core instructional material. Students were aware AI hallucinates, citing concerns about missing or incorrect information. This skepticism aligns with student attitudes towards text-based LLMs, where distrust and hesitancy are common \cite{zastudil_generative_2023, amoozadeh_trust_2024, pratherEtAl-2023-copilotNoviceProgrammers}. Students also echoed concerns that AI content may interfere with deeper learning \cite{bouvier_rest_2025, rogers_attitudes_2024}. Notably, many students did not find AI videos appropriate for complex topics. This is particularly concerning for novice CS students, who are more likely to over-rely on inaccurate AI outputs without critically questioning them \cite{hou_effects_2024, zastudil_generative_2023, bouvier_rest_2025, prather_robots_2023}. 

This distrust appears strongest when students imagine AI videos replacing, rather than supplementing, instructor-led content.  Students worried AI videos would be less engaging, with its dull delivery, and would diminish valuable interactions with their instructors. This is consistent with prior work noting that students value human connection over AI presenters \cite{vallis_student_2024, weerakoon_enhancing_2024} and that GenAI cannot replace human instructors more broadly \cite{prather_robots_2023}. Hou et al. similarly reported that GenAI tools increased CS students' feelings of isolation \cite{houEtA-2025-genAIErodeSocial}, though their concern was on student-student rather than student-instructor relationships which our participants sought to protect. Students also trusted instructors' domain expertise over AI, appreciating how instructors draw on personal and industry experience to make content relatable. They also raised concerns about AI making education less worth its cost and AI's societal impact generally, echoing CS instructors' views in Lau and Guo's study \cite{lauAndGao-2023-instructorsAdapt}. However, these concerns were tied to wholesale replacement rather than supplemental use. In all, AI videos can play a meaningful role in computing education without displacing instructors' presence that students value.

%% file: 7_limitations.tex
\section{Limitations}
Several limitations should be considered when interpreting the findings of this exploratory descriptive study. First, the study was designed to describe students' perceptions of AI videos and not to determine whether the videos were more effective than other instructional materials. Nevertheless, students scored an average of 86\% on a post-survey Markdown quiz, suggesting learning from these videos. However, this reflects only short-term knowledge acquisition, not retention or transfer. Future work should examine other learning outcomes using controlled experimental designs.

Second, the findings may be influenced by students’ prior CS background and by the brief duration of the instructional videos. Although most students reported little or no prior knowledge of Markdown language, they were enrolled in higher-level CS courses and likely had experience with programming languages and technical notation more complex than the content presented in the videos. This prior experience may have made the material easier to learn and may limit generalizability to learners with less computing experience. In addition, more complex content may not be well suited to a short-form video format and require extended explanation. Future work should examine students’ perceptions of AI videos in foundational programming courses such as CS1 and CS2. Future work should also explore longer or more varied formats, such as live coding and program demonstrations, for more complex content.

Third, our qualitative analysis may limit generalizability and is open to multiple interpretations. To address these limitations, we transparently describe our coding process and include representative quotes to strengthen the reliability and credibility of the findings and their scope of interpretation.

% Wilson - combining below two as we have less space
% Second, the findings may be influenced by the students' prior computing background. Although most of the students reported having little or no prior knowledge of GitHub's markup language, they were enrolled in higher-level computing courses and likely had experience with programming languages and technical notation more complex than the markup language addressed in the videos. This background may have made the topic easier to learn and may limit generalizability to students with less computing experience or to topics that require more substantial conceptual development. 

% Third, the length of the videos may also limit the generalizability of the findings. Each video was approximately three minutes long and addressed a focused topic. Although this format may have supported student engagement and comprehension, many computing topics require longer explanations, extended examples, or opportunities for practice. Therefore, students' perceptions and quiz performance in this study may not generalize to longer AI-generated videos or to more complex topics that cannot be effectively introduced in brief instructional segments.

%% file: 8_conclusion.tex
\section{Conclusion}
In this study, we investigated how students perceive AI-generated videos in CS education. Through a post-survey study with 170 students, we found while students viewed AI videos on Markdown as high quality and usable, they remained hesitant about their broader adoption in courses. Students favored AI videos for simple, supplemental, and visual topics, but expressed concerns about content quality, reduced instructor interaction, and diminished educational value. 
% As AI-generated videos continue to improve, and as trust in this technology develops through advances in video generation models and instructor-in-the-loop review processes, student perceptions and acceptance of these videos may evolve. 
Ultimately, we hope our work empowers CS instructors to adopt AI videos in ways that best support student learning.

\begin{acks}
    We thank Nicholas Ho for their initial work, and Knowlify for providing early access to its platform and complimentary video credits for this study. Knowlify had no role in the study design, data collection, analysis, interpretation, or the decision to publish.
\end{acks}